\documentclass[aps,pre,twocolumn,nofootinbib,showpacs]{revtex4}
\usepackage{amsmath}
\usepackage{amssymb}
\usepackage{epsfig}

\def\S{{\cal S}}
\def\A{{\cal A}}
\def\B{{\cal B}}

\def\be{\begin{equation}}
\def\en{\end{equation}}

\def\RR{\rm \hbox{I\kern-.2em\hbox{R}}}
\def\NN{\rm \hbox{I\kern-.2em\hbox{N}}}
\def\ZZ{\rm {{Z}\kern-.28em{Z}}}
\def\CC{\rm \hbox{C\kern -.5em {\raise .32ex \hbox{$\scriptscriptstyle
|$}}\kern
-.22em{\raise .6ex \hbox{$\scriptscriptstyle |$}}\kern .4em}}

\def\<{\langle}
\def\>{\rangle}

\begin{document}
\title{Multifractal stationary random measures and 
multifractal random walks with log-infinitely divisible scaling laws}

\author{Jean-Fran\c{c}ois Muzy}
\affiliation{CNRS UMR 6134, Universit\'e de Corse, Quartier Grossetti,
20250, Corte, France}
\email{muzy@univ-corse.fr}
\author{Emmanuel Bacry}
\affiliation{Centre de Math\'ematiques appliqu\'ees, Ecole Polytechnique,
91128 Palaiseau, France}
\email{emmanuel.bacry@polytechnique.fr}

\date{June 2, 2002}

\begin{abstract}
We define a large
class of continuous time multifractal random measures and processes with 
arbitrary log-infinitely divisible exact or asymptotic scaling law.
These processes generalize within a unified framework both the recently
defined log-normal Multifractal Random Walk (MRW) \cite{mrwepj,mrwpre} 
and the log-Poisson ``product of cynlindrical pulses" \cite{barral}.
Our construction is based on some ``continuous
stochastic multiplication'' (as introduced in \cite{schmitt}) 
from coarse to fine scales that
can be seen as a continuous interpolation of discrete
multiplicative cascades. 
We prove the stochastic convergence of the defined processes and
study their main statistical properties.
The question of genericity (universality) of limit
multifractal processes is addressed within
this new framework. We finally provide a method
for numerical simulations and discuss some specific examples.
\end{abstract}


\maketitle

\section{Introduction}
Multifractal processes are now widely used models in many areas including
nonlinear physics, geophysics or econophysics. They are used to
account for scale-invariance properties of some observed data. 
Our purpose in this article is to introduce a wide class
of random measures and stochastic processes with stationary increments
that possess exact multifractal
scaling without {\em any preferred scale ratio}.
This construction generalizes (and unifies) 
the recently introduced log-normal ``Multifractal Random Walk'' model
\cite{mrwepj,mrwpre} and the log-Poisson compounded ``Multifractal
Product of Cynlindrical Pulses'' \cite{barral}. Technical mathematical proofs are reported in the companion paper \cite{mrwjsp}.

In the
multifractal framework, scale-invariance properties of a 1d
stochastic
process $X(t)$
are generally characterized by the exponents
$\zeta_q$ which govern the power law scaling of the absolute moments of the
``fluctuation'' $\delta_l X(t)$ of $X(t)$ at any scale $l$ up to a scale
$T$, i.e.,
\begin{equation}
\label{escaling}
m(q,l) = K_q  l^{\zeta_q}, ~\forall l \le T,
\end{equation}

where $m(q,l)$ is defined as the expectation (from now on,
the symbol $E(.)$ will always refer to the mathematical expectation):
\begin{equation}
\label{moments}
m(q,l) = E\left[|\delta_l X(t)|^q\right].
\end{equation}

The so-called ``fluctuation'' $\delta_l X(t)$ can be defined in various
ways, the most commonly used definition being
\begin{equation}
\label{increment}
\delta_l X(t) = X(t+l)-X(t).
\end{equation}
When  the exponent $\zeta_q$ is linear in $q$, i.e., $\exists H$, $\zeta_q =
qH$, the process is referred to as a {\em monofractal} process. Let us note
that the so-called {\em self-similar} processes (e.g., fractionnal Brownian
motions, Levy walks) are a particular case of
monofractal processes. On the contrary, if $\zeta_q$ is a non-linear
function of $q$ it is referred to as a {\em multifractal} process
(or as a process displaying {\em multiscaling} or {\em intermittency}).

The scale-invariance property (\ref{escaling}) can be qualified as a
{\em continuous} scale-invariance in the sense that the relation is a
strict equality for the continuum  of scales  $0< l \le T$. Alternative
``weaker'' forms of 
scale-invariance have been widely used. All of them
assume  at least the asymptotic scale-invariance relation
\begin{equation}
\label{ascaling}
m(q,l) \sim K_q  l^{\zeta_q}, ~l \rightarrow 0^+.
\end{equation}
The {\em discrete} scale-invariance property adds strict equality for
discrete scale values $l_n$ (with $l_n \rightarrow 0^+$, when $n \rightarrow
+\infty$)
\begin{equation}
\label{dscaling}
m(q,l_n) = K_q  l_n^{\zeta_q}, ~n \rightarrow +\infty.
\end{equation}

The image implicitely associated with a multifractal process is a random
multiplicative cascade from coarse to fine scales. Such cascading processes
can be explicitely defined in very different ways ranging from the
original construction proposed by Mandelbrot in its early work \cite{man74} to
recent wavelet-based variants \cite{jmp}. 
Though these different constructions
do not lead to the same objects, they all aim at building a stochastic
process $X(t)$ which fluctuation $\delta_{\lambda l} X(t)$ at a scale $\lambda l$ (where $l$ is an arbitrary scale
smaller than the large scale $T$ and $\lambda < 1$) is obtained from its
fluctuation $\delta_{l} X(t)$ at the larger scale $l$
through the simple ``cascading'' rule
 \begin{equation}
\label{cascade}
\delta_{\lambda l} X(t) \operatornamewithlimits{=}^{law} W_\lambda \delta_l X(t),~\forall l\le T,
\end{equation}
where $W_\lambda$ is a positive random variable independent of $X$ and which
law depends only on $\lambda$.

In the multiplicative construction by Mandelbrot \cite{man74}
(which log-normal variant was originally introduced by Yaglom \cite{Yaglom}),  
a positive multifractal
measure $M(dt)$ is built, i.e., $X(t) = M([0,t])$ is an increasing process
and the fluctuation  $\delta_l X(t)$ at scale $l$ is defined by
Eq.~(\ref{increment}).
In the more recent wavelet-based constructions $X(t)$ is a process, not
necessarily increasing, and the fluctuation $\delta_l X(t)$  is defined as
the wavelet coefficient at scale $l$ and time $t$ \cite{jmp}.
Such cascade processes have been extensively 
used for modeling scale-invariance
properties as expressed in Eq.~(\ref{escaling}).
However, apart from self-similar processes which are monofractal processes
with continuous dilation-invariance properties,
all cascade processes involve some arbitrary discrete scale ratio and
consequently only have discrete scale-invariance
properties (Eq.~(\ref{dscaling})).   
Indeed, they rely on  
a ``coarse to fine'' approach consisting first in fixing the fluctuations at
the large scale $T$ and then, using Eq.~(\ref{cascade}) iteratively, deriving
fluctuations at smaller and smaller scales.
However, the $(t,l)$ time-scale half-plane  is very constrained : one cannot
choose freely
$\delta_l X(t)$ at all scales and times. Thus, the cascade is generally not
built using the whole time-scale half-plane
but only  using a sparse (e.g., dyadic) grid in this half-plane.
Consequently an arbitrary scale ratio (e.g., $\lambda = 1/2$) is introduced
in the construction.  The continuous scale-invariance property
(\ref{escaling}) is thus broken and replaced by the weaker discrete
scale-invariance property (\ref{dscaling}) with  $l_n = \lambda^nT$.
Moreover, as a consequence, 
the fluctuations $\delta_l X(t)$ in these approaches 
are no longer stationary.

Let us point out that despite the potential interest
of stochastic  
processes with stationary fluctuations
and continuous invariance-scaling properties, 
until recently \cite{mrwepj,mrwpre, barral}, explicit constructions possessing such properties
were lacking. Let us assume that one can build a continuous cascade process $X(t)$ 
satisfying Eq.~(\ref{cascade}) with a continuous dilation 
parameter $\lambda$.
As first pointed out by Novikov \cite{novikov}, 
if such a construction is possible, a simple transitivity argument shows that
$\ln W_\lambda$ must have an infinitely
divisible law: its characteristic function is of the form $\hat
G_\lambda(q) = \lambda^{F(q)}$. 
One can then easily prove that
$
m(q,\lambda l) = m(q,l) \lambda^{F(-iq)},
$ 
and consequently $\zeta_q = F(-iq)$.
Continuous cascade statistics and log-infinite divisibility
have been the subject of many works with applications in
various domains ranging from turbulence to 
geophysics \cite{novikov,saito,novikovb,casdub,sheway,
marsan96,sche97,aArn97,aArn98c}.
In the case $W_\lambda = \lambda^H$ is deterministic, one gets
$\zeta_q = qH$ and therefore $X(t)$ is a monofractal process. This is the
case  of the so-called self-similar processes. The simplest non-linear
(i.e., multifractal) case is the so-called log-normal cascade that
corresponds to a log-normal law for $W_\lambda$ and thus to a parabolic
$\zeta_q$ spectrum. Other well known log-infinitely divisible
models often used in the context of fully developed
turbulence are log-Poisson \cite{sl} and log-Levy \cite{scherlev}
models.

From our
knowledge, among all the attempts to build multifractal processes with
continuous scale invariance properties and stationary fluctuations, 
only the  recent works by
Bacry et al. \cite{mrwepj, mrwpre} and Barral and Mandelbrot \cite{barral} 
refer to a precise mathematical construction.
Bacry et al. have built the so-called Multifractal Random Walk (MRW) processes
as limit processes based on discrete time random walks
with stochastic log-normal variance.
Independently, Barral and Mandelbrot \cite{barral} have proposed a new class
of stationary multifractal measures. Their construction is
based on distributing, in a half-plane, Poisson points associated with 
independent identically distributed (i.i.d.)
random weights and then taking a product of these weights over conical domains.
In the same time, Schmitt and Marsan considered an extension
of discrete cascades to a continuous scale framework \cite{schmitt}
and introduced infinitely divisible stochastic intregrals 
over cone-like structures similar to those of Barral and Mandelbrot.
However they did not considered any continuous time limit
of their construction nor studied its scaling properties as
defined by Eq.~(\ref{escaling}).

In this paper, we propose a model that is based on the stochastic 
approach developed in \cite{schmitt} and that unifies all previous 
constructions within a single framework.
Starting from original discrete multiplicative cascades,
we will use a cone-like construction as in \cite{schmitt} and
\cite{barral} in order to get rid of discrete scale ratios 
and to consider any
log-infinitely divisible multifractal statistics. 
We will show that this
allows us to build a very large class of multifractal measures
and processes (including
original MRW \cite{mrwepj,mrwpre} and Barral-Mandelbrot 
multifractal measures \cite{barral}) for
which both 
long range correlations and multiscaling properties can be controlled very
easily. 

The paper is organized as follows.
In section II, we review the 
discrete multiplicative cascades in order to naturally introduce
the notion of stochastic integral over a cone-like
structure in some ``time-scale'' half-plane.
We then 
define a class of log-infinitely divisible 
stationary Multifractal Random Measures (MRM)
which statistical properties are studied in section III.
In section IV, we define the log-infinitely
divisible Multifractal Random Walks. 
We show how their scaling properties can
be inferred from the associated MRM.
In section V we address some questions
related to numerical simulations and provide
explicit examples.
In section VI, we discuss some links of the present
work with previous connected approaches.
Conclusions and prospects for future research 
are reported in section VII.

\section {From discrete multiplicative measures to multifractal
random measures}

In this section, we provide some heuristics about how one can build a
positive
stationary stochastic measure $M(dt)$ with continuous scale-invariance
properties, i.e., such that the associated increasing process $X(t) =
M([0,t])$ satisfies Eq.~(\ref{cascade}) with a continuous dilation
parameter $\lambda$.

\subsection {Discrete multiplicative cascades }
For the sake of illustration, 
let us start with simple discrete multiplicative
cascades.
In the original construction, as proposed by Mandelbrot,
one builds the measure $M(dt)$ as the
limit of a sequence of stochastic measures $M_{l_n}$,
indexed by a discrete scale
parameter $l_n = T\lambda^n$ (we choose  $\lambda=1/2$). 
The measure $M_{l_n}$, at  the step $n$ of the construction,
has a uniform density on successive dyadic
intervals  of size $l_n$.
The idea is to build the sequence $M_{l_n}$ so that it statisfies (when $n$ is
varying) a scale-invariance property. The $n$-th measure
is obtained from the $(n-1)$-th measure by multiplication
with a positive random process $W$ which law does not depend on $n$. 
One can naturally index the dyadic intervals along the dyadic
tree using a kneading sequence $\{s_1 \ldots s_n\}$ where
$s_i=0$ (resp. $s_i=1$) if, at ``depth''
$i$, the interval is on the left (resp. on the right) boundary
of its parent interval. 
Thus, for instance, one gets the following dyadic intervals: 
$I_{0} = [0,2^{-1}]$, $I_{1} = [2^{-1},1]$,
$I_{00}= [0,2^{-2}]$, $I_{01}=[2^{-2},2^{-1}]$ and so on...
With these notations, the multiplicative rule reads (see
Fig. 1(a)): 
\begin{equation}
\label{iterM}
  M_{l_n}(I_{s_0 \ldots s_n}) = W_{s_0 \ldots s_n} M_{l_{n-1}}
(I_{s_0 \ldots s_{n-1}})
\end{equation}
where $I_{s_0 \ldots s_n}$ is of size $T 2^{-n}$ and is one of the two ``sons'' of the interval
$I_{s_0 \ldots s_{n-1}}$ and $W_{s_0 \ldots s_n}$ are 
i.i.d.
Since the construction is invariant
with respect to a rescaling by a factor $2$,
the limit measure $M(dt)$ will then
satisfy the same scale-invariance property and
will be multifractal in the discrete sense of Eq.~(\ref{dscaling}).
There is a huge mathematical literature devoted to the
study of such a construction
and we refer the reader to Refs. \cite{kp,dl,guiv,molchan} for
rigourous results about the existence, regularity
and statistical properties of Mandelbrot cascades.
In physics or other applied sciences, as recalled in the
introduction, the previous construction
(and many of its variants) is considered as the paradigm
for multifractal objects and has been often used
as a reference model in order to reproduce observed
multiscaling.
But because of its lack of continuous scale invariance
and translation invariance, such models cannot 
be fully statisfactory in many contexts
were the considered phenomena possess
some degree of stationarity and do not display
any preferred scale ratio.

\begin{figure}
\begin{center}
\includegraphics[width=8cm]{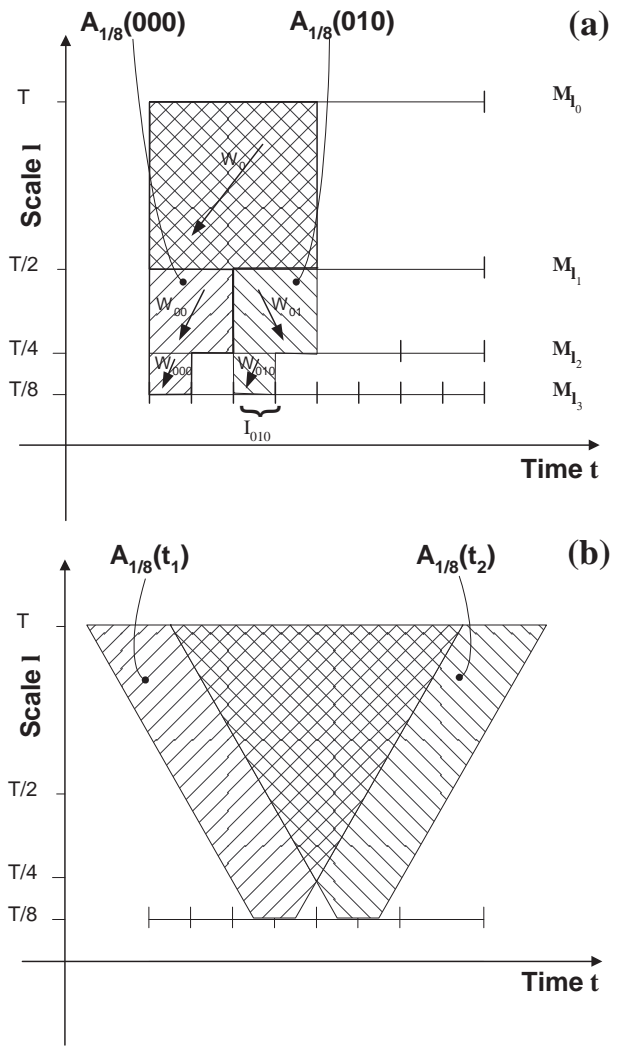}
\end{center}
\caption{Defining a continuous cascade interpolating Mandelbrot cascades.
(a) Mandelbrot cascade: One starts from the coarse scale $T$ and
constructs the sequence $M_{l_n}$ that is resolved at scale $T 2^{-n}$
by iterating a multiplicative rule along the dyadic tree.
At each construction step,
the random multiplicative factor $W_{s_0 \ldots s_n}$ can be represented
as the exponential of the integral
of an infinitely divisible noise over a square domain indicated
by hatched sets.
(b) Continuous cascade conical domains obtained as linear smoothing
of the hatched domains in (a). Such cones are involved in the definition
of the process $\omega_{2^{-3}}(t)$}
\end{figure}

\subsection{Revisiting discrete cascades}
\label{revisiting}
In order to generalize  Eq.~(\ref{iterM}) to a continuous
framework one can try to perform the limit $\lambda \rightarrow 1$
in the discrete construction as in Ref. \cite{schmitt}.
Another way to proceed
is to represent a Mandelbrot cascade as a discretization
of an underlying continuous construction.

For that purpose, we suppose that
the random weights $W_{s_0 \ldots s_n}$ in the cascade
are log-infinitely divisible.
For a continuous cascade, this choice 
can be motivated as follows:
Let us suppose that the large scale density, 
$M_T(dt)$ is equal (or proportional) to 
the Lebesgue measure $dt$.
We would like to
define iteratively the densities
$M_{l}(t)$, for all $l\le T$.
To go from resolution $l'$ to resolution $l \leq l'$, we thus write
\begin{equation}
\label{cc}
M_{l}(t) \operatornamewithlimits{=}^{law} W_{l/l'}(t) M_{l'}(t) ,~\forall l \le l',
\end{equation}
where $W_{l/l'}(t)$ is a positive stationary discrete random process
independent of $M_l'(t)$.
Let us define 
\begin{equation}
\omega_{l}(t) =  \ln W_{l/T}(t).
\end{equation}
Thus, one gets, 
\begin{equation}
M_l(t)dt = e^{\omega_{l}(t)} M_T(t)dt = e^{\omega_{l}(t)} dt.
\end{equation}
By iterating Eq.~(\ref{cc}), we see that 
$W_{l/T}(t)$ can be
obtained as a ``continuous
product'' of positive i.i.d. random variables. Consequently
$\omega_l(t)$ can be written as the sum of an arbitrary number
of i.i.d. random variables. This is precisely the definition of
an infinitely divisible random variable \cite{Feller}.
In order to go from discrete to continuous cascades,
it is therefore natural to assume that, in the discrete situation, 
$\omega_{l_n}(s_0 \ldots s_n) = \ln\left(\prod_{i=0}^n W_{s_0...s_i} \right)$
is infinitely divisible.
A ``simple'' way
of building such an infinitely divisible process
$\omega_{l_n}$ is to represent it by a stochastic integral
of an infinitely divisible
stochastic 2D measure $P(dt,dl)$
over a domain $\A_{l_n}(s_0 \ldots s_n)$:
\begin{equation}
 \omega_{l_n}(s_0 \ldots s_n) = P(\A_{l_n}(s_0 \ldots s_n)) =
 \int_{\A_{l_n}(s_0 \ldots s_n)} \! \! \! \! \! \! \! \! P(dt,dl) 
\end{equation}
The stochastic measure $P(dt,dl)$ is
uniformly (with respect to a measure
$\mu(dt,dl)$)
distributed  on the time-scale
half-plane $\S^+ =  \{(t,l),~t \in \mathbb{R} , \; l \in \mathbb{R}^+\}$.

Let us recall that, by definition\footnote{we refer the reader to ref. \cite{rajput}
for a rigourous definition of the so-called ``infinitely divisible independently scattered random measure''},
if $P(dt,dl)$ is a stochastic infinitely divisible measure
uniformly (with respect to a
measure $\mu(dt,dl)$)
distributed on $\S^+$,  then,
for any two $\mu$-mesurable sets $\A$ and $\A'$ such that $\mu(\A)=\mu(\A')$,
then $P(\A)$ and $P(\A')$ are identically distributed 
random variables which characteristic
function is nothing but
\begin{equation}
\label{escaling1}
 E \left( e^{ipP(\A)} \right) = e^{\varphi(p)\mu(\A)},
\end{equation}
where $\varphi(p)$ 
depends only on a centering parameter $m$
and the so-called canonical L\'evy measure $\nu(dx)$
which is associated with $P$. The general
shape of $\varphi$ is described
by the celebrated L\'evy-Khintchine formula \cite{Feller,Bertoin}:
\begin{equation}
\label{LK}
  \varphi(p) = imp +\int \frac{e^{ipx}-1-ip \sin x}{x^2} \nu(dx).
\end{equation}
with $\int_{-\infty}^{-y} \nu(dx)/x^2 < \infty $ and 
$\int_{y}^{\infty} \nu(dx)/x^2 < \infty $ for all $y > 0$.

The sets $\A_{l_n}(s_0 \ldots s_n)$ 
associated with each of the $2^{n}$ values
$\omega_{l_n=T2^{-n}}(s_0 \ldots s_n)$ in Mandelbrot
construction
can be chosen naturally as the union of all similar squares
$I_{s_0 \ldots s_k} \times [T 2^{-k}, T 2^{-(k-1)}]$, $n \leq k \leq 1$:
\begin{equation}
\label{domdis}
  \A_{l_n}(s_0 \ldots s_n) = \cup_{k=0}^{n} I_{s_0 \ldots s_k} \times 
[T 2^{-k}, T 2^{-(k-1)}] \; .
\end{equation}
The sets $\A_{2^{-3}}(000)$ and $\A_{2^{-3}}(010)$ 
are indicated as hatched domains in Fig. 1(a).
Since the $W_{s_0 \ldots s_n}$'s are i.i.d., we want to choose the measure $\mu(dt,dl)$
such that 
the measure of each square $I_{s_0 \ldots s_n} \times [T 2^{-n}, T 2^{-(n-1)}]$
is a constant. The natural measure to choose is $\mu(dt,dl) = dtdl/l^2$. It
is the natural measure associated with the
time-scale plane $\S^+$ in the sense that it is (left-) invariant by the
translation-dilation group.

Fixing $\mu(dt,dl) = dtdl/l^2$ one then gets
$\forall n$,
\[
\int_{I_{s_0 \ldots s_n} \times [T 2^{-n}, T 2^{-(n-1)}]} \! \! \! \! \! \! \! \! \! \!
d\mu = \int_{2^{-n}}^{2^{-(n-1)}} l^{-2} dl \int_0^{2^{-n}} dt = 1/2
\]
and thus, from Eq.~(\ref{escaling1}), $W_{s_0 \ldots s_n}$ are
i.i.d.: we recover exactly the Mandelbrot construction.

\subsection {Towards multiplicative cascades  with continuous
scale-invariance}
We would like to apply the previous scheme in the case
the construction is no longer indexed by 
a discrete scale parameter $l_n$ but by a continuous
scale parameter $l$. 

In order to ``interpolate'' smoothly
this construction both in time and scale,
on can interpolate the previous union of
similar squares  $\A_{l_n}(s_0 \ldots s_n)$ 
(Eq.~(\ref{domdis})) using domains  
$\A_{l}(t)$ where $(t,l)$ can take any value in $\S^+$.  In order the limit measure $M(dt) = \lim_{l\rightarrow 0} M_l(dt)$ to be stationary, it is clear that one has  to choose the set $\A_{l}(t)$ to be 
``translation-invariant'' in the sense that
\begin{equation}
(t+\tau,l') \in \A_{l}(t+\tau) \Longleftrightarrow (t,l') \in
\A_{l}(t),~~ \forall \tau.
\end{equation}
One thus just needs to specify the set $\A_{l}(0)$. 
A ``natural'' choice 
(though, as we will see in section \ref{asymptscaling}, 
it only leads to asymptotic, and not exact, scaling properties) 
seems to be the conical set (see Fig. 1(b))
\begin{equation}
\label{Al0}
A_{l}(0) = \{(t,l'), l' \geq l, -f(l')/2 <  t \le f(l')/2 \},
\end{equation}
where $f(l)$ is the function 
\begin{equation}
\label{fpasexact}
  f(l) = \left\{
\begin{array}{ll}
 & l \; \mbox{for} \; l \leq T \\
 & 0 \; \mbox{for} \; l \geq T,
\end{array}
\right.
\end{equation}
In case case, we get exactly the construction originally proposed,
using a densification argument, by Schmitt and Marsan in Ref. \cite{schmitt}.
Other choices for the function $f(l)$ 
will be discussed in section \ref{asymptscaling}.

Let us remark that the choice of the linear conical shape $f(l)=l$
is consistent with the interpretation of the parameter $l$ as a scale
parameter. Indeed, the value
of the measure $P$ around some position $(t_0,l)$ in the half-plane $\S^+$
influences the values of the measure over the time
interval $]t_0-l/2,t_0+l/2]$, i.e., exactly over a time scale $l$.

\section{Stationary Multifractal Random Measures}
\subsection{Defining MRM}
\label{secdef}
According to the arguments of the previous section, we thus 
propose the following definition for the
class of log-infinitely divisible 
multifractal random measures.
Let us 
introduce an infinitely
divisible stochastic 2D measure $P$ uniformly distributed on the half-plane
$\S^+ =  \{(t,l),~t\in \mathbb{R},~l \in \mathbb{R}^+\}$ with respect to 
the measure $\mu(dt,dl) =
dtdl/l^2$ and associated with the Levy measure
$\nu(dx)$. 
Let us recall that for any set $\A \subset \S^+$, 
$P(\A)$ has an infinitely divisible law whose
moment generating function is
\begin{equation}
\label{mmm}
 E \left( e^{pP(\A)} \right) = e^{\varphi(-ip)\mu(\A)},
\end{equation}
where $\varphi(p)$ is defined by Eq.~(\ref{LK}).
Henceforth, we define the real convex cumulant generating 
function, as, when it exists, 
\begin{equation}
\label{defpsi}
   \psi(p) \equiv \varphi(-ip) \; .
\end{equation} 
Let $\omega_l(t)$ the stationary stochastic process defined by
\begin{equation}
\label{defomega}
\omega_l(t) = P\left(\A_l(t)\right),
\end{equation}
where $\A_l(t)$ is the 2D subset of $\S^+$ defined by
\begin{equation}
\label{Al}
A_{l}(t) = \{(t',l'), ~ l'\ge l,~-f(l')/2 < t'-t \le f(l')/2\},
\end{equation}
where $f(l)$ satisfies:
\begin{equation}
\label{fexact}
  f(l) = \left\{
\begin{array}{ll}
 & l \; \mbox{for} \; l \leq T \\
 & T \; \mbox{for} \; l \geq T,
\end{array}
\right.
\end{equation}
As it will be shown in the following sections,
the choice of the large scale behavior $f(l)= T$ for $l \geq T$
is the unique one that ensures the convergence of the
construction and the exact scaling of the limit measure.
The cone-like domain $\A_l(t)$ is indicated as an hatched domain
in Fig. 2.
\begin{figure}[t]
\begin{center}
\includegraphics[width=8cm]{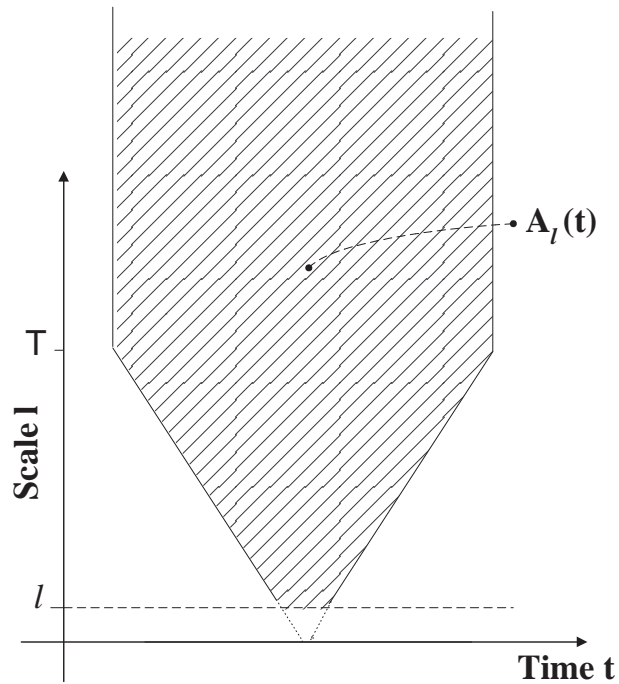}
\end{center}
\caption{Conical domain (Eq.~(\ref{fexact})) in the $(t,l)$ half-plane involved
in the definition of $\omega_l$ (Eq.~(\ref{defomega})).}
\end{figure}

We finally define the stochastic positive measure $M_l(dt)$ as
\begin{equation}
\label{Ml}
M_l(dt) = e^{\omega_l(t)} dt,
\end{equation}
meaning that for any Lebesgue measurable set $I$, one has\footnote{Since
the paths of $\omega_l(t)$ are continuous to the right
and limited to the left, the integral (\ref{defml}) is well defined (see \cite{mrwjsp})}
\begin{equation}
\label{defml}
M_l(I) = \int_I e^{\omega_l(t)} dt.
\end{equation}

The MRM $M$ is then obtained as the limit measure (the meaning and
the existence of this
limit will be addressed in the next section)
\begin{equation}
M(dt) = \lim_{l \rightarrow 0^+} M_l(dt).
\end{equation}
Since a simple change in the mean of the stochastic measure $P$ would lead
to the same measure up to a deterministic multiplicative factor, we will assume, without
loss of generality, that $\psi$ satisfies
\begin{equation}
\psi(1) = 0.
\end{equation}

We can consider some generalizations of the previous construction. The
first one consists in changing
the function $f(l)$, i.e., to change the
shape $\A_l$ the measure $P$ is integrated on. The second one consists
in changing the measure $\mu(dt,dl)$, i.e., to change the way the measure
$P$
is distributed  in the half-plane $\S^+$.
Actually, from Eq.~(\ref{escaling1}) one can easily show that the construction
only depends on the function $\mu(\A_l)$, consequently, changing the
shape of 
$f(l)$ basically amounts changing $\mu(dt,dl)$.
A simple example that illustrates such a freedom is the choice 
$\mu(dt,dl) = dt dl$, i.e. $\mu$ is nothing but 
the 2D Lebesgue measure.
In that case, $\mu(A_l)$ remains unchanged if one chooses
$f(l) = 1/l$ in the definition (\ref{fexact}).
The parameter $l$ is no longer a scale but can
be interpreted as a frequency and thus $\S^+$ is the
time-frequency half-plane.
Therefore, 
in the following sections, without loss of generality, we choose to fix
$\mu(dt,dl) = dtdl/l^2$ (i.e. to work within the time-scale half-plane $\S^+$) 
and we will discuss, in section \ref{asymptscaling}, the consequences which rise
from other choices than (\ref{fexact}) for the function $f(l)$.

\subsection{Existence of the limit MRM $M(dt)$}
In \cite{mrwjsp}, we prove, within
the framework of positive continuous martingales that,
almost surely, $M_l$  converges to a well defined limit
measure when $l\rightarrow 0^+$.
Moreover, we prove that if $\psi'(1) < 1$ then there exists
$\epsilon > 0$ such that $\psi(1+\epsilon) < 1$ and the moment of order $1+\epsilon$,
$E \left( M_l([0,t])^{1+\epsilon}\right)$,
is finite. Then, using the fact that $\psi(1)=0$,
it is straightforward to prove that
\begin{equation}
E\left(M([0,t]\right) = \lim_{l\rightarrow 0^+} E\left(M_l([0,t]\right) = 1.
\end{equation}
Consequently the limit measure 
$M(dt)$ is non
degenerated (i.e., different from zero).
The overall proof is very technical and, for this reason, has not been reproduced in this paper. 

However, if $\psi(2)<1$, one can prove (see Appendix A) that
$\sup_l E(M_l[0,t]^2)$ is bounded and 
that the sequence $M_l(dt)$ converges in the mean square sense.
Again, using the fact that  $\psi(1)=0$, it follows that $M(dt)$ is non degenerated.
(Let us note that, as explained  in section III D, one can prove  \cite{mrwjsp} that assuming 
$\psi(2)<1$ basically amounts assuming that $E\left(M([0,t])^2\right) < +\infty$).

\subsection{Exact multifractal scaling of $M(dt)$}
In order to study the scaling properties of the limit measure
$M(dt)$, let us establish the scale invariance properties
of the process $\omega_l(t)$.

\subsubsection{Characteristic function of  $\omega_l(t)$}
Let $q\in \mathbb{N}^*$, ${\bf t_{q}} = {t_1,t_2,\ldots,t_q}$ with $t_1 \le t_2 \le
\ldots \le t_n$ and
${\bf p_{q}} = {p_1,p_2,\ldots,p_q}$. The characteristic function of the
vector $\{\omega_l(t_m)\}_{1\le m\le q}$ is defined by
\begin{equation} 
\label{charomega}
Q_l({\bf t_{q}}, {\bf p_{q}}) = E \left( e^{\sum_{m=1}^q i p_m P(\A_{l}(t_m))}\right).
\end{equation}
Relation (\ref{mmm}) allows us to get an expression for
quantities like
$E \left( e^{\sum_{m=1}^q a_m P(\B_m)} \right)$ where $\{\B_m\}_m$ would be disjoint
subsets of $\S^+$ and $a_m$ arbitrary numbers. However the
$\{\A_{l}(t_m)\}_m$ in Eq. (\ref{charomega}) 
have no reason to be disjoint subsets. We need to find a
decomposition of $\{\A_{l}(t_m)\}_m$ onto disjoint domains. This is
naturally done by considering the different intersections between these
domains.

Let us define the cone intersection domains as:
\begin{equation}
\A_l(t,t') = \A_l(t)\cap\A_l(t').
\end{equation}
and 
\begin{equation}
\label{defrho}
\rho_l(t) = \mu \left( \A_l(0,t) \right) \; .
\end{equation}
Using the definition of $\A_l(t)$ with the shape
of $f(l)$ as given by Eq.~(\ref{fexact}), the
expression for $\rho_l(t)$ reads:
\begin{equation}
\label{rhoexact}
\rho_{l}(t) =
\left\{
\begin{array}{ll} 
  \ln \left(\frac{T}{l} \right)+1-\frac{t}{l} & \mbox{if}~t \leq l \\
\ln \left(\frac{T}{t} \right) & \mbox{if}~T \ge t\ge l \\
0 & \mbox{if}~ t > T
\end{array}
\right.,
\end{equation} 
Notice that $\rho_l(t)$ satisfies the remarkable property (for $t \leq T$ and
$\lambda \leq 1$):
\begin{equation}
  \rho_{\lambda l}(\lambda t) = \rho_l(t)-\ln(\lambda) .
\end{equation}
 
In Ref. \cite{mrwjsp}, $Q_l({\bf t_{q}} ,{\bf p_{q}} )$ is computed
using a recurrence on $q$.
We obtain the following result:
\begin{equation} 
\label{char}
Q_l({\bf t_{q}} ,{\bf p_{q}} )  = e^{\sum_{j=1}^q \sum_{k=1}^{j}
\alpha(j,k) \rho_l(t_k-t_j)},
\end{equation}
where
\begin{equation} 
\label{defalpha}
\alpha(j,k) = \varphi(r_{k,j})+\varphi(r_{k+1,j-1})-\varphi(r_{k,j-1})-\varphi(r_{k+1,j}),
\end{equation}
and
\begin{equation}
r_{k,j} = 
\left\{
\begin{array}{lll}
& \sum_{m=k}^j p_m, & {\mbox{for}}~k \le j \\
& 0,  & {\mbox{for}} ~k > j
\end{array}
\right.
\end{equation}
Moreover, let us remark that
\begin{equation}
\label{rk}
\sum_{j=1}^q \sum_{k=1}^{j}\alpha(j,k) = \varphi \left( \sum_{k=1}^q p_k \right).
\end{equation}

\subsubsection{Multifractal properties of $M(dt)$}
The multifractal scaling properties of the limit MRM $M(dt)$ result
from the scale invariance
property of the process $\omega_l(t)$ that is itself
a direct consequence of 
previous exponential expression (\ref{char}) for
the characteristic function of the process $\omega_l$ and
the particular shape of the conical domains
leading to expression (\ref{rhoexact}).
Indeed, using these equations together with Eq. (\ref{rk}), 
it can be proven that
$\forall n$, $\forall t_1,\ldots t_n \in [0,T]^n$, $\forall p_1 \ldots
p_n \in \mathbb{R}^n$, one has
\[
 Q_{\lambda l}(\lambda {\bf t_{q}} ,{\bf p_{q}} )= \lambda^{-\varphi(\sum_{j=1}^n
 p_j)} Q_l({\bf t_{q}} ,{\bf p_{q}} ) \; .
\] 
It follows that, for $\lambda \leq 1$, 
the process $\omega_l(t)$ satisfies, for $l \in [0,T]$, the 
following invariance property:
\begin{equation}
  \{ \omega_{\lambda l}(t \lambda) \}_t \operatornamewithlimits{=}^{law} \{
  \Omega_\lambda + \omega_l(t) \}_t
\end{equation}
where $\Omega_\lambda$ is an infinitely divisible  random variable (ie., it does not depend on $t$) which is independent of $\omega_l(t)$ and
which infinitely divisible law is defined by:
\begin{equation}
  E(e^{ip\Omega_\lambda}) = \lambda^{-\varphi(p)}
\end{equation}

We deduce the following scale invariance relationship 
for the sequence of measures $M_l([0,t])$:
\begin{eqnarray}
\nonumber
& & \{M_{\lambda l} ([0,\lambda t])\}_t  = \{\int_0^{\lambda t}
e^{\omega_{\lambda l}(u)}du\}_t \\
\nonumber
& & \lambda \{\int_0^{t} e^{\omega_{\lambda l}(\lambda u)}du\}_t 
\operatornamewithlimits{=}^{law}  \lambda e^{\Omega_{\lambda}} \{\int_0^{t}
e^{\omega_{l}(u)}du\}_t \\ 
& & =  \lambda e^{\Omega_{\lambda}} \{M_{l} ([0,t])\}_t 
=  W_{\lambda}  \{M_{l}([0,t])\}_t,
\end{eqnarray}
where $W_\lambda = \lambda e^{\Omega_\lambda}$ 
is independent of $\{M_{\lambda l} ([0,\lambda t])\}_t$.

By taking the limit $l \rightarrow 0^+$, one gets the 
continuous cascade equation for MRM as  
defined in Eq.~(\ref{cascade}):
\begin{equation}
\label{scalingomega}
\{M([0,\lambda t])\}_t \operatornamewithlimits{=}^{law} W_{\lambda}
\{M([0, t])\}_t,~~~\forall \lambda \in ]0,1],
\end{equation}
where $\ln(W_\lambda)$ is an infinitely divisible random variable
independent of $\{M([0, t])\}_t$.

The exact multifractal scaling follows immediately:
$\forall q \in \mathbb{R}, ~\forall t< T,~~$, we get
\begin{equation}
\label{mescaling}
E\left[M([0,t])^{q}\right] = K_q t^{\zeta_q},
\end{equation}
were the multifractal spectrum and the prefactor read:
\begin{eqnarray}
\label{zetaid}
\zeta_q & = & q-\psi(q) \\
\label{prefid}
K_q & = & T^{-\zeta_q} E \left[ M([0,T])^{q}\right] 
\end{eqnarray}
Let us notice that the moment $E\left(M([0,t])^{q}\right)$ in 
Eq.~(\ref{mescaling})
can be infinite. Conditions for $E\left(M([0,t])^{q}\right)$ to be finite will be discussed in section III D.

\subsubsection{Multi-scaling of correlation functions}
The exact (multi-)scaling law (\ref{mescaling}) for the  
absolute moments can be easily extended,  along the same line,
to $n$-points correlation functions.
Indeed, let us define the n-point correlation
function, when it exists, as
\begin{multline}
 C_n(t_1,\ldots,t_n;\tau_1,\ldots,\tau_n;p_1, \ldots,p_n) = \\ 
 E \left[ M([t_1,t_1+\tau_1])^{p_1} \ldots M([t_n,t_n+\tau_n])^{p_n} \right] .
\end{multline}
where $\tau_k > 0$, and $t_k + \tau_k \leq t_{k+1}$.
It is easy to show that if $t_n+\tau_n-t_1 < T$, then
$C_n$ is an homogeneous function of degree $\zeta_{\sum p_k}$: 
\begin{equation}
\label{scalingcf}
  C_n(\{\lambda t_k\};\{\lambda \tau_k \}; \{p_k\}) =  \lambda^{\zeta_{p}}
  C_n(\{t_k\};\{\tau_k \}; \{p_k\}) 
\end{equation}
with
\begin{equation}
p = \sum_{k=1}^n p_k \; .
\end{equation}
This equation extends the scaling law (\ref{mescaling}) for the moments
to multi-points correlation functions.

\subsection{Algebraic tails of probability density functions}
\label{tail}
Let us note that, if $q > 1$, 
\begin{eqnarray*} 
E\left[M([0,t])^q\right] & = & E\left[ \left(M([0,t/2]) +
    M([t/2,t])\right)^q\right] \\ & \ge &  2 E \left[ M([0,t/2])^q \right]. 
\end{eqnarray*}
Using the multifractal scaling (\ref{mescaling}),
one gets $l^{\zeta_q} \ge 2 (l/2)^{\zeta_q}$ leading to $1 \ge
2^{1-\zeta_q}$. It follows that 
\begin{equation}
\label{mom1}
K_q < \infty \Rightarrow \zeta_q \ge 1.
\end{equation} 
Thus, if $q> 1$, the  $q$ order moment is infinite
if $\zeta_q < 1$. In  \cite{mrwjsp}, we show that the
``reverse'' implication is also true, i.e.,
\begin{equation}
\label{mom2}
\zeta_q > 1 \Rightarrow K_q < \infty.
\end{equation}
Let us notice that this infinite moment condition $\zeta_q < 1$, $q > 1$, 
is exactly the same as for discrete multiplicative cascades
established in Refs. \cite{kp,dl,guiv}. Divergence of moments for
multifractals have also been discussed in e.g. \cite{man74,sche97}.

Therefore, if there exists some value $1 < q_\ast < \infty$ such that
\begin{equation}
\label{infmom}
   \zeta_{q_\ast} = 1
\end{equation}
then, 
\begin{equation}
   \mbox{Prob}\left\{M([0,t]) \geq x \right\} \sim x^{-q_\ast}, \; \mbox{when} \; x \rightarrow +\infty
\end{equation}
The pdf of the measure $M([0,t])$ is thus heavy tailed
with a tail exponent that can be, unlike classical $\alpha$-stable
laws, arbitrary large.

\subsection{Examples}
\label{examples}
In order to illustrate previous considerations
let us consider some specific examples.
\begin{itemize}
\item {\em Deterministic case:} \\
The simplest situation is when the Levy $\nu(dx)$ 
measure is identically zero: 
This case corresponds
to the self-similar, monofractal situation where $\psi(p) = pm$.
The constraint $\psi(1) = 0$ implies $m=0$, and we thus get the
Lebesgue measure.

\item {\em Log-normal MRM:}\\ 
The log-normal MRM is obtained when
the canonical measures attributes a finite mass at the origin: 
$\nu(dx) = \lambda^2 \delta(x) dx$ and $\lambda^2 > 0$.
From Eqs. (\ref{LK}) and (\ref{defpsi}), the cumulant
generating function is that of a normal distribution:
$\psi(p) = pm +\lambda^2 p^2/2$.
The condition $\psi(1)=0$ implies the relationship $m=-\lambda^2/2$.
The log-normal $\zeta_q$ spectrum is a parabola:
\begin{equation}
\label{lnzeta}
 \zeta_q^{\mbox{ln}} = q(1+\frac{\lambda^2}{2}) -\frac{\lambda^2}{2}q^2.
\end{equation}
Let us note that the so-obtained increasing process $M([0,t])$ is the same as the increasing MRW process mentioned in the conclusion of  \cite{mrwpre}. This similarity will be further  discussed in  section VI.A.1.

\item {\em Log-Poisson MRM:} \\ 
When there is a finite mass at some finite value $x_0= \ln(\delta)$,
of intensity $\lambda^2 = \gamma (\ln \delta)^2$: 
$\nu(dx) = \lambda^2 \delta(x-x_0)$.
The corresponding distribution is Poisson of scale parameter $\gamma$ and
intensity $\ln(\delta)$:
$\psi(p) = p\left(m-\sin(\ln(\delta))\right) -\gamma(1-\delta^p)$.
The log-Poisson $\zeta_q$ spectrum is therefore exactly the
same as the one proposed by She and L\'ev\^eque in
their cascade model for turbulence \cite{sl}:
\begin{equation}
\label{lpzeta}
  \zeta_q^{\mbox{lp}} = q m' + \gamma(1-\delta^q) 
\end{equation}
where $m'$ is such that $\zeta_1 = 1$.
Notice that in the limit $\delta \rightarrow 1^-$, $\gamma (\ln \delta)^2 \rightarrow
\lambda^2$, one recovers the log-normal situation.
In the original She-L\'ev\^eque model, $\gamma = 2$ and
$\delta = 2/3$ \cite{sl}.

\item {\em Log-Poisson compound MRM:} \\
When the canonical measure $\nu(dx)$ satisfies 
$\int \nu(dx) x^{-2} = C < \infty$
(e.g. $\nu(dx)$ is concentrated
away from the origin),
one can see that $F(dx) = \nu(dx)x^{-2}/C$ is a probability measure. 
In that case, 
\[
 \varphi(p) = im'p + C \int (e^{ipx}-1) F(dx)
\]
is exactly the cumulant generating function 
associated with a poisson process with
scale $C$ and compound with the distribution $F$ \cite{Feller}.
Let us now consider a random variable $W$ such that
$\ln W$ is distributed according to $F(dx)$.
It is easy to see that $\int e^{px} F(dx)= E(W^p)$.
It turns out that the log-Poisson compound MRM
has the following multifractal spectrum:
\begin{equation}
\label{lpczeta}
  \zeta_q^{\mbox{lpc}} = qm-C\left(E(W^q)-1\right)
\end{equation}
This is exactly the spectrum obtained by
Barral and Mandelbrot in their construction
of ``product of cylindrical pulses'' \cite{barral}. 
The similarity between our construction and Barral and Mandelbrot construction will be further discussed in  section VI.A.2.

\item {\em Log-$\alpha$ stable MRM:} \\
When $\nu(dx) \sim x^{1 -\alpha} dx$ for $0 < \alpha < 2$, 
one has an log $\alpha$-stable
MRM: 
\begin{equation} 
\label{lszeta}
 \zeta_q^{\mbox{ls}} = qm-\sigma^{\alpha} |q|^{\alpha}
\end{equation}
Such laws have been used in the context of 
turbulence and geophysics \cite{sche97,scherlev}.
They have been often referred to as ``universal multifractals''
because $\alpha$-stable laws are fixed points of infinitely
divisible laws under a suitable renormalization procedure.
 
\end{itemize}
Many other families of $\zeta_q$ spectra can be obtained
(e.g. log-Gamma, log-Hyperbolic,...) for other
choices of the Levy measure.
Let us remark that in the case of a normal random variable $\omega_l$,
the function $\rho_l(\tau)$ is nothing but the covariance
of the process as introduced in Refs. \cite{mrwepj,mrwpre}.
This function, that measures the areas of domains ${\cal A}_l$ intersections,
is therefore the analog of the covariance
for general infinitely divisible distributions.
The equation (\ref{char}) shows that our construction
can be seen as a natural extension of Gaussian processes
(or multivariate Gaussian laws) within the class of infinitely
divisible processes (multivariate infinitely divisible laws)
in the sense that it is completely characterized
by a cumulant generating functions $\varphi(p)$ (specifying
the mean and the variance parameters in the Gaussian case) 
and a 2-points ``covariance" function $\rho(t)$ (or covariance
matrix).

\subsection{Asymptotic scaling and universality}
\label{asymptscaling}
In the last section, we have seen that the choice (\ref{fexact}) for $f(l)$ leads to exact scaling of the moments of the associated MRM. 
In this section we study the scaling behavior of the moments
for other choices of $f(l)$
\footnote{We recall that we do not consider 
other choices for the uniform measure $\mu(dt,dl)$ 
in the half space $\S^+$ because
$f(l)$ and $\mu(dt,dl)$ are involved in the properties of
the limit measure $M(dt)$ only through the
function $\rho_l(t) = \mu(A_l(0,t))$. Hence,
up to a change of variable $l' = h(l)$, one
can always set $\mu(dt,dl) = dt dl /l^{-2}$.}.
Let us remark that $f(l)$ is defined
up to a multiplicative constant.
This just amounts to a choice in the scale
of $\psi(p)$. \\In the following,  $f^{(e)}(l)$ will refer to the ``exact scaling'' choice (\ref{fexact}) that was made in the previous sections. The so-obtained sets in the $\S^+$ half-plane will be referred to as $\A^{(e)}_l(t)$, the $\omega$ process as $\omega_l^{(e)}(t)$ and the associated MRM as $M^{(e)}(dt)$. 

\subsubsection{Large scale perturbation of $f(l)$}\noindent
Rigorous mathematical proofs can be found in \cite{mrwjsp}. \\
Let us first study the case when one builds an MRM using a function $f(l)$ which differs from $f^{(e)}(l)$ only for scales larger than a large (fixed) scale $L$, i.e.,
\begin{equation}
\label{flargescale}
  f(l) = \left\{
\begin{array}{ll}
 & f^{(e)}(l) \; \mbox{for} \; l \leq L \\
 & g(l) \; \mbox{for} \; l \ge L,
\end{array}
\right.
\end{equation}

For the measure $M_l(dt)$ to remain finite, $\rho_l(t)$ must be
finite and thus we must have
$\int_l^{+\infty} f(u)  u^{-2}du < \infty $. 
Therefore,
the large scale behavior $g(l)$ must be such that,
for some $\epsilon > 0$,
$g(l) = O(l^{1-\epsilon})$ as $l \rightarrow +\infty$.
An example of such large scale modification is the
function defined by (\ref{fpasexact}), where $g= g^{(s)}$ where $g^{(s)}(l)=0$
and $L=T$. This function is the one which was used by Barral  and Mandelbrot
in Ref. \cite{barral}.
Let us first choose the particular case $g = g^{(s)}$ ($L$ being any strictly positive number). 
The so-obtained sets in the $\S^+$ half-plane will be referred to as $\A^{(s)}_l(t)$, the $\omega$ process as $\omega^{(s)}_l(t)$ and the associated MRM as $M^{(s)}(dt)$.
Since, $\forall$ $t,t'$, $(\A^{(e)}_l(t) \backslash \A^{(s)}_l(t)) \cap \A^{(s)}_l(t')
= \emptyset$, one has
\[
 \omega^{(e)}_l(t) = \omega_l^{(s)}(t) + \delta_L(t)
\]
where $\delta_L(t) = P(\A^{(e)}_l(t) \backslash \A^{(s)}_l(t))$ is a process which is independent of the process $\omega^{(s)}_l(t)$ and which does not depend on the value of $l$ (as long as $l<L$). It follows that
\begin{eqnarray*}
 E\left( M^{(e)}([0,t])^q \right) & \leq & E\left(\sup_{[0,t]}e^{q
 \delta_L(t)}\right)E\left( M^{(s)}([0,t])^q \right) \\
 E\left( M^{(e)}([0,t])^q \right) & \geq & E\left(\inf_{[0,t]}e^{q \delta_L(t)}\right)E\left( M^{(s)}([0,t])^q \right)
\end{eqnarray*}
Because the process $\delta_L(t)$ is (right) continuous,
$\lim_{t \rightarrow 0^+} \sup_{[0,t]}e^{q  \delta_L(t)} = \lim_{t \rightarrow
  0^+} \inf_{[0,t]}e^{q\delta_L(t)} = e^{q \delta_L(0)}$, we get 
\begin{equation}
E\left( M^{(s)}([0,t])^q \right) \operatornamewithlimits{\sim}_{t \rightarrow 0} C_q E\left( M^{(e)}([0,t])^q \right).
\end{equation}
Thanks to the exact scaling
of $M^{(e)}([0,t])$ (Eq.~(\ref{mescaling})), we see
that, 
\begin{equation}
E\left( M^{(s)}([0,t])^q \right) \operatornamewithlimits{\sim}_{t \rightarrow 0} C_q t^{\zeta_q}
\end{equation}
where $\zeta_q$ is defined in Eq.~(\ref{zetaid}).Thus, we find that $M^{(s)}(dt)$, corresponding to the specific choice $g(l) = g^{(s)}(l) = 0$, satisfies the asymptotic scale invariance property (\ref{ascaling}).
If one chooses a different function $g(l)$, using exactly the same arguments as above (in which $M^{(e)}(dt)$ is replaced by $M(dt)$, i.e., the measure obtained when using the new function $g(l)$), one can prove \cite{mrwjsp} that 
\begin{equation}
E\left( M([0,t])^q \right) \operatornamewithlimits{\sim}_{t \rightarrow 0} D_q E\left( M^{(s)}([0,t])^q \right),
\end{equation}
and consequently
\begin{equation}
E\left( M([0,t])^q \right) \operatornamewithlimits{\sim}_{t \rightarrow 0} D_q C_q t^{\zeta_q}.
\end{equation}
Thus, using any function $g(l)$ in (\ref{flargescale})  (satisfying $g(l) = O(l^{1-\epsilon})$ ($l \rightarrow +\infty$))
leads to an MRM measure which satisfies the asymptotic scale invariance property (\ref{ascaling}).

\subsubsection{Small scale perturbation of $f(l)$}
Let us now study the consequences of a small scale
perturbation of $f^{(e)}(l)$.
Let us
suppose that $f(l) \sim l^{\alpha}$ for $l \rightarrow 0$.
In that case,
\be
\rho_l(0) \operatornamewithlimits{\sim}_{l \rightarrow 0} \left\{
\begin{array}{ll} 
  l^{\alpha-1} & \; \mbox{if} \; \alpha < 1 \\
  Cst & \; \mbox{if} \; \alpha > 1 \\
  -\ln(l) & \; \mbox{if} \;  \alpha = 1
\end{array}
\right.,
\end{equation}
In the first case, $\alpha < 1$, 
one can show any moment of order $1+\epsilon$ ($\epsilon > 0)$ 
cannot be bounded. Since $E(M_l([0,t])) = 1$, general martingale
arguments can be used to prove that $M_l([0,t])$ converges
towards the trivial zero measure.

If $\alpha > 1$, the limit measure is proportional
to the Lebesgue measure and thus one obtains
the trivial asymptotic scaling:
\[
  E(M([0,t])^q) \operatornamewithlimits{\sim}_{t \rightarrow 0} C_q t^q
\]

Let us now consider the marginal case, $f(l)= l+o(l)$ at small scales.
From the results of previous section, with no loss of
generality, as far as the asymptotic scaling is concerned,
we can suppose that the large scale component
is identical to the exact scaling situation: $f(l) = T$ for $l \geq T$.
We can then show that, 
once again, the MRM satisfies a multifractal scaling 
in the asymptotic sense
(Eq.~\ref{ascaling}):
\be
\label{as}
  E(M([0,t])^q) \operatornamewithlimits{\sim}_{t \rightarrow 0} K'_q t^{\zeta_q}
\end{equation}
Indeed, let us index all
the quantities by the cut-off scale $l$
and the integral scale $T$, i.e., we add an explicit reference to the
integral scale $T$:  
$\rho_{l,T}(t)$ is the area of domain
intersections and $M_{l,T}([0,t])$ is the
associated MRM.
After some little algebra, using the definition
of $\rho_{l,T}(t)$ and the fact that $f(l) = l + o(l)$, 
one can show that 
\begin{equation}
  \rho_{\lambda l, \lambda T}(\lambda t)  
\operatornamewithlimits{\longrightarrow}_{\lambda \rightarrow 0}
  \rho^{(e)}_{l,T}(t) \; .
\end{equation}
Thus, thanks to Eq.~(\ref{char}), we have 
\[
\{ \omega_{\lambda l,\lambda T}(\lambda t) \}_t 
 \operatornamewithlimits{\longrightarrow}^{law}_{\lambda \rightarrow 0}
 \{ \omega_{l,T}^{(e)}(t) \}_t
\] 
Hence, because $M_{l,T}([0,t]) = \int_0^t e^{\omega_{l,T}(u)} du$,
by taking the limit $l \rightarrow 0$,
\[
\lambda^{-1} M_{0, \lambda T}([0,\lambda t])
\operatornamewithlimits{\longrightarrow}^{law}_{\lambda \rightarrow 0} 
M^{(e)}_{0,T}([0,t])
\]
and therefore, from Eqs. (\ref{mescaling}), (\ref{zetaid}) and
(\ref{prefid}),  
\[
 \lambda^{-q} E\left( M_{0, \lambda T}([0,\lambda t])^q \right) \operatornamewithlimits{\longrightarrow}_{\lambda
 \rightarrow 0}  E\left(M^{(e)}_{0,T}([0,T])^q\right) T^{-\zeta_q} t^{\zeta_q} 
\]
By choosing $T = T' \lambda^{-1}$ and using the identity
$M^{(e)}_{0,\lambda^{-1} T'}([0,\lambda^{-1} T']) =^{law} \lambda^{-1} 
M^{(e)}_{0,T'}([0,T'])$, we conclude that
\[
 E\left( M_{0,T'}([0,\lambda t])^q \right) \operatornamewithlimits{\sim}_{\lambda
 \rightarrow 0}  \lambda^{\zeta_q}   
\]
This achieves the proof.

We can therefore see, that, as far as asymptotic multifractality
is concerned, the pertinent parameter is the small scale 
behavior of the function $f(l)$ or equivalently
the small time behavior of $\rho_{l=0}(t)$.
As pointed out previously, in the case of a Gaussian field $\omega(t)$
(i.e. the infinitely divisible law has only a Gaussian component),
$\rho(t)$ is nothing but the covariance of the process.
The previous discussion leads thus to the conclusion
that {\em non trivial limit multifractal measures arise
only in the marginal situation when the correlation
function of the logarithm of the fluctuations decreases
as a logarithmic function.}

\subsection{An alternative discrete time construction for MRM}
\label{secdmrsm}

In the case $\int x^{-2} \nu(dx) < \infty$ (e.g. the Levy measure 
has no mass in an interval around
$x=0$) a realization of the measure $P(dt,dl)$ is made of dirac
functions distributed in the $\S^+$ half-plane.
Thus the process $M_l(([0,t]) = \int_0^t e^{\omega_l(t)} dt$ is a jump
process that can be simulated with no approximation.

However, if  $x^{-2} \nu(dx)$ has a non finite integral 
(e.g., it has a Gaussian component), this is no longer the case.
Thus, one has to build another sequence of stochastic measures $\tilde
M_l(dt)$ that converges in law towards $M(dt)$ and that can be seen
as a discretized version of $M(dt)$.
We will see, in following sections, that such a discrete time approach
is also interesting for multifractal stochastic processes construction.

We choose $\tilde M_l(dt)$ to be uniform on each interval
$[kl,(k+1)l[$, $\forall k \in \mathbb{N}$ and with density $e^{\omega_l(kl)}$.
Thus, for any $t > 0$ such that $t = pl$ with $p \in \mathbb{N}^*$, one gets
\be
\label{tildeM}
\tilde M_l([0,t]) = \sum_{k=0}^{p-1} e^{\omega_l(kl)} l.
\end{equation}
In the same way as for the measure $M_l(dt)$, one can prove \cite{mrwjsp},
within the framework of positive martingales, that, almost surely, $ \tilde
M_{l=2^{-n}}(dt)$ converges towards a well defined limit measure when $n
\rightarrow +\infty$ (i.e., $l \rightarrow 0^+$).
Moreover, in the same way as in section III B, if we suppose $\psi(2)<1$,
then one can show (the proof is very similar as the one in Appendix A) that,
in the mean square sense,
\be
\lim_{n\rightarrow +\infty}  \tilde M_{l=2^{-n}}(dt) = M(dt).
\end{equation}
As long as $\psi(2)<1$, this construction gives therefore a way of
generating a measure which is arbitrary close (by choosing $l$ small enough)
to the limit measure $M(dt)$.

\section{Log-infinitely divisible multifractal random walks}
In this section we build and study
a class of multifractal stochastic processes
that are no longer, as before,
strictly increasing processes (measures).
They can be basically built in two different ways: (i) By subordinating
a fractional Brownian motion (fBm) with the previously defined
MRM $M(t)$ or (ii) by a stochastic integration of a MRM against
a fractional Gaussian noise (fGn).
As we will see, most of the statistical properties
of the so-obtained random processes
are directly inherited from those of the associated MRM.
 
\subsection{MRW with uncorrelated increments}
In the same spirit as the
log-normal MRW constructions in Refs. \cite{mrwepj,mrwpre}
(see also \cite{seuront}), we use a stochastic
integration of $e^{\omega_l(t)}$ against the (independent) Wiener measure
$dW(t)$.

\subsubsection{Definition}
Since $E(e^{\omega_l(t)}) = 1$, one can consider the process
\be
\label{defmrw}
  X_l(t) = \int_0^t e^{\frac{1}{2} \omega_l(u)} dW(u)
\end{equation}
where $dW(t)$ is a Gaussian white noise independent of $\omega_l$.
The MRW is then defined as the limit of $X_l(t)$ when $l \rightarrow 0^+$ :
\be
X(t) = \lim_{l \rightarrow 0^+} X_l(t).
\end{equation}
One can easily prove that for fixed $t$ and $l$, one has
\be
\label{convlaw}
  X_l(t) \operatornamewithlimits{=}^{law} \sigma_l(t) \epsilon
\end{equation} 
where $\epsilon$ is a standardized normal random variable
independent of $\sigma_l(t)$ which is itself nothing
but the associated MRM as defined previously:
\be
\label{defsv}
  \sigma_l^2(t) = \int_0^t e^{\omega_l(u)} du = M_l(t) \; .
\end{equation}
Let us note that the (non decreasing) increments of $\sigma^2_l(t)$
is referred to (in the field of mathematical finance \cite{shi,bnqf}) as the
{\em stochastic volatility}.

Using the same kind of arguments on finite dimensional laws, one can also
prove that 
the finite dimensional laws of the process $X_l(t)$ converge to those of the
subordinated process $B(M(t))$. Actually, one can show \cite{mrwjsp} that,
as long as $\psi'(1)<1$ ( i.e., $\exists \epsilon>0$, $\psi(1+\epsilon)<1$
which is, as mentioned in section \ref{revisiting},  
the condition for the limit
measure $M(dt)$ to be non degenerated), on has
\be
\label{yesyes}
X(t) =   \lim_{l\rightarrow 0^+} X_l(t)  \operatornamewithlimits{=}^{law}
\lim_{l\rightarrow 0^+} B(\sigma_l(t)) \operatornamewithlimits{=}^{law}
B(M(t)).
\end{equation}

The so-obtained MRW can be thus understood as a Brownian motion in a
``multifractal time'' $M(t)$.
The subordination of a Brownian process with a non decreasing
process has been introduced by Mandelbrot and Taylor \cite{aMan67}
and is the subject of an extensive literature in mathematical finance.
Multifractal subordinators have been considered
by Mandelbrot and co-workers\cite{aMan99} and widely used
to build multifractal processes from multifractal measures (see below).
In a forthcoming section we will see that multifractal
subordination and stochastic integration do not lead to
the same processes when one considers long range correlated
Gaussian noises (fGn).

\subsubsection{Expression of the moments and multifractal properties}
Thanks to Eq.~(\ref{yesyes}) (assuming $\psi'(1)<1$), one gets the
expression of 
the absolute moments of $X(t)$ (or $X(t_0+t)-X(t_0)$):
\begin{eqnarray*}
\forall q,~~ E\left(|X(t)|^q\right) & = & E(|\epsilon|^{q})
E\left(M([0,t])^{q/2}\right) \\
 & = & \sigma^q \frac{2^{q/2} \Gamma(\frac{q+1}{2})}{\Gamma(\frac{1}{2})}
E\left(M([0,t])^{q/2}\right) \; ,
\end{eqnarray*}
where the first factor comes the order $q$ moment of a centered Gaussian
variable of variance $\sigma^2$.

If $M(dt)$ is an exact multifractal stationary random measure,
then, $X(t)$ obeys the exact multifractal scaling equation:
\be
E \left( |X(t)|^q \right) = \sigma^q \frac{2^{q/2}
  \Gamma(\frac{q+1}{2})}{\Gamma(\frac{1}{2})} K_{q/2} t^{\zeta_q}
\end{equation}
where $K_q$ is defined in Eq. (\ref{prefid}) and
\be
\label{zetaXu}
  \zeta_q = q/2-\psi(q/2) \; .
\end{equation}
Using  Eqs. (\ref{mom1}), (\ref{mom2}) and (\ref{infmom}), one 
deduces that, for $q > 2$,
\be
E \left( |X(t)|^q \right)  < +\infty \Rightarrow \zeta_q \ge 1,
\end{equation}
and conversely,
\be
 \zeta_q < 1  \Rightarrow  E \left( |X(t)|^q \right) < \infty.
\end{equation}
Moreover, let us note that if $M(dt)$ verifies only an asymptotic scaling,
so does the
MRW process $X(t)$.

\subsubsection{An alternative discrete time construction for MRW processes}

As in Refs. \cite{mrwepj,mrwpre} or in
section \ref{secdmrsm}, one can also try to build an MRW
process using a discrete approach.
A discrete construction can be useful for numerical simulations.
Let us, for instance, choose $l_n = 2^{-n}$. And let  $\epsilon_l[k] =
\int_{kl}^{(k+1)l} dW(u)$ be a discrete
Gaussian white noise.
We define the piece-wise constant process $\tilde X_{l_n}(t)$ as
($t = pl_n$):
\be
\tilde X_{l_n}(t) = \sum_{k=0}^{p-1} e^{\frac{1}{2} \omega_{l_n}(kl_n)}
\epsilon_{l_n}[k]\; .
\end{equation}
The MRW $\tilde X_{l_n}(t)$ can be rewritten as
\be
\tilde X_{l_n}(t) = \sum_{k=0}^{p-1} \sqrt{\tilde M_{l_n}([kl_n,(k+1)l_n])}
\epsilon_{l_n}[k]/l_n\; ,
\end{equation}
where $\tilde M_{l_n}$ is defined by (\ref{tildeM}).
One then deduces easily the convergence of $\tilde X_{l_n}$ from the
convergence of 
$\tilde M_{l_n}$. Thus, for instance, one can prove \cite{mrwjsp} that, as long as $\psi(2)<1$, one has
\be
 \lim_{n\rightarrow +\infty} \tilde X_{l_n}(t)  \operatornamewithlimits{=}^{law} B(M(t)).
\end{equation}

\subsection{MRW processes with long-range correlations}
\label{longrangemrw}
\subsubsection{Definitions}
In order to construct long-range correlated MRW, it is
natural to replace the Wiener noise (resp. Brownian motion)
in previous construction
by a fractional Gaussian noise (resp. fractional Brownian motion).
A fBm, $B_H(t)$ is a continuous, self-similar, zero-mean 
Gaussian process which 
covariance reads (see e.g. \cite{Taqqu} for a precise
definition and properties):
\be
  E\left( B_H(t) B_H(s) \right) = \frac{\sigma^2}{2} \left(s^{2H}+t^{2H}-|t-s|^{2H} \right)
\end{equation}
where $0<H<1$ is often called the Hurst parameter.
Standard Brownian motion corresponds to $H=1/2$.

The simplest approach to construct a long-range correlated 
MRW follows the idea of Mandelbrot
that simply consists in subordinating a fractional
Brownian motion of index $H$ with the
MRM $M(t)$, i.e.,
\begin{equation}
  X_H^s(t) = B_H \left[ M(t)\right] \; . 
\end{equation}

An alternative would consist in buidling a stochastic integral
against a fGn $dW_H(t)$
\be
\label{xfgn}
  X_{l,H}^i(t) = \int_0^t e^{\omega_l(t)} dW_H(t) 
\end{equation}
and considering some appropriate limit $l \rightarrow 0$.
However stochastic integrals against fGn cannot be
defined as easily as for the white Gaussian noise and
the proposed constructions require the complex machinery 
of Malliavin calculus or Wick products\cite{Taqqu2,Decr}.
One simple way to define the previous integral could
be to see it as the limit of a Riemann sum: 
\be
\label{mrwfmb}
\int_0^t e^{\omega_l(t)} dW_H(t) \equiv \lim_{\Delta t \rightarrow 0} 
\sum_{k=0}^{t/\Delta t} e^{\omega_{l}(k \Delta t)} \epsilon_{H,\Delta t}[k] \; \,
\end{equation} 
where $\epsilon_{H,\Delta t}(k) = B_H(k\Delta t)-B_H((k-1)\Delta t)$.
We have not proved yet that this is a mathematically sound
definition.
However, if one assumes that (\ref{xfgn}) makes sense,
one can address the question of the existence of the
limit process $\lim_{l \rightarrow 0} X_{l,H}^i(t)$.
In Appendix B, we provide heuristic arguments for 
mean square convergence.
We obtain a condition:
\begin{equation}
\label{condi}
  H > 1/2 + \psi(2)/2
\end{equation}
where $\psi(p)$ is the cumulant generating function associated
with $\omega$.

\subsubsection{Multifractal properties}
In the case of the 
subordinated version $X^s_H(t)$ of the MRW, the 
scaling properties can be directly deduced by the self-similarity
of $B_H(t)$ \cite{aMan99}.
Since $B_H(t+\tau)-B_H(t) \operatornamewithlimits{=}^{law} \tau^H (B_H(t+1)-B_H(t))$,
and $M(t)$ is independent of $B_H$, one has
$X_H(M(t)) \operatornamewithlimits{=}^{law} M(t)^H B_H(1)$. The scaling of the
absolute moments of the increments of $X_H$ is therefore:
\begin{eqnarray*}
 E\left[ |X_H(t)|^q \right] & = & E \left[ M(t)^{qH} \right] E \left[|B(1)|^q
 \right] \\
 & = & K_{qH} G_q t^{\zeta^s_q}  
\end{eqnarray*}
with
\be
   \zeta^s_q = qH-\psi(qH)
\end{equation}

For the second version, $X_H^i$, the scaling of the moments is
determined using the scale-invariance of the process $\omega_l(t)$
and the self-similarity of the fGn $dW_H(t)$. Using the same method as for the measure
$M(t)$, one obtains:
\be
 E\left[ |X_H^i(t)|^q \right] = M_q t^{\zeta^i_q} 
\end{equation}
with
\be
   \zeta^i_q = qH-\psi(q)
\end{equation}
We can see that the multifractal spectra of $X_H^s$
and $X_H^i$ are different: they do not correspond to identical
processes as it was the case for the uncorrelated construction.
Notice that the existence criterion (\ref{condi}), can be simply
rewritten as $\zeta^i_2 > 1$. According to the considerations
developed in section \ref{tail}, this condition ensures the
existence of the second order moment of $X_H^i$.
Wether the class of processes $X_H^i(t)$ can be 
extended, in some weak probabilistic sense, 
to values of $H < 1/2 + \psi(2)/2$ is still
an open problem (such processes would have an infinite variance).  
The condition of finite variance for $X_H^s(t)$ is less
restrictive since it comes to the condition 
$K_{2H} < \infty$, where $K_{2H}$ is defined in Eq.~(\ref{prefid}).
For $H < 1/2$ such a moment is always finite.

\subsection{A remark on subordination}
Let us remark that, in some sense, the subordination
by a MRM $M(dt)$ can be iterated.
Indeed, if $M_1([0,t])$ and $M_2([0,t])$ are two independent MRM, 
the subordinated measure 
\be
\label{subor} 
M([t_1,t_2]) = M_1\left(M_2([0,t_1]),M_2([0,t_2])]\right)
\end{equation}
is well defined.
Using the cascade equation (\ref{scalingomega}),
we deduce that 
\begin{eqnarray}
\label{toto}
 M([0,\lambda t]) & \operatornamewithlimits{=}^{law} &
 M_1\left(0,W_{\lambda}^{(2)}M_2([0,t]\right) \\
& \operatornamewithlimits{=}^{law} & W^{(1)}_{W^{(2)}_\lambda} M([0,t])
\label{subormes}
\end{eqnarray}
where $W_{\lambda}^{(1,2)}$ are the (independent) 
log-infinitely divisible weights associated with the MRM $M_{1,2}$.
The second equality is valid only if $W_{\lambda}^{(2)} < 1$,
i.e., when the Levy measure associated with $M_2$ is 
concentrated on $]0, +\infty]$.
By computing the moment of order $q$ of both sides of the
equality, we see that the mulifractal spectrum
of the subordinated measure reads:
\be
\label{subspect}
  \zeta_q = \zeta^{(2)}_{\zeta^{(1)}_q} = q-\psi^{(2)}\left(q-\psi^{(1)}(q)\right)
\end{equation} 
where $\zeta^{(i)}_q$ (resp. $\psi^{(i)}(q)$), $i=1,2$, is the spectrum associated with $M_i$.
The equation (\ref{subormes}) corresponds to a ``randomization" of
the rescaling factor $\lambda$ that parametrizes
the log-infinitely divisible law of $\ln(W^{(1)}_\lambda)$.
It is easy to prove \cite{Feller} that the law
of $\ln(W^{(1)}_{W^{(2)}_\lambda})$ remains infinetely
divisible. The class of log-infinitely divisible MRM
is therefore closed under subordination.
The family of subordinated spectra (\ref{subspect})
is thus included in the family of log-infinitely divisible
spectra and the operation (\ref{subor}) does not allow
us to build new MRM with exact scale invariance properties.

\section{Numerical simulations}
\subsection{Principles}
In order to generate realizations of $\tilde M_{l_n}(dt)$ 
(defined by (\ref{tildeM})), one needs to be
able to generate realizations of $\omega_{l_n}(k l_n)$. However, following
the definition of $\omega_l$ one would need
realizations of the 2d random measure $P(dt,dl)$ for $l\ge l_n$.
In the case of a compound poisson process,
the process $M_l([0,t]) = \int_0^t e^{\omega_l(t)} dt$ is a jump
process that can be synthetized easily.  

In the general case, we need to find a set of disjoint ``elementary'' domains
of the half-plane $\S^+$  such that,
for any $k$, there exists a subset of this set  such that
$\A_{l_n}(kl_n)$ can be expressed as a the union over the elementary domains
of this subset.
Since, at fixed $l_n$, the boundaries of the domains $\A_{l_n}(kl_n)$ ($k \in
\ZZ$) define a tiling of $\S^+$, it is natural to consider
the elementary cells of this tiling. Each cell is the intersection
between left and right strips limited by left and right boundaries
of conical domains:  
Let us define the cell ${\cal B}_{l}(t,t')$ (with $t<t'$)
as 
\be
{\cal B}_{l}(t,t') = \left({\cal A}_{l}(t)\backslash{\cal A}_l(t-l)\right) \cap
\left({\cal A}_{l}(t')\backslash{\cal A}_l(t'+l)\right)
\end{equation}
Then, by definition the cells $\{{\cal B}_{l_n}(kl_n,k'l_n)\}_{k<k'}$ are
disjoint domains and form a partition of the subspace (of $\S^+$) $\{(t,l)
\in \S^+,~~ l\ge l_n\}$.
Moreover
\be
\A_{l_n}(kl_n) = \bigcup_{-\infty\le i \le k} ~ \bigcup_{k\le j \le +\infty}
{\cal B}_{l_n}(il_n,jl_n).
\end{equation}
On the other hand,
for a fixed $s\ge l_n$, one has
\[
(u,s) \in \A_{l}(t)\backslash\A_l(t-l) \Leftrightarrow t-l +f(s)/2\le u
\le t+f(s)/2,
\]
and
\[
(u,s) \in \A_{l}(t+l)\backslash\A_l(t) \Leftrightarrow t -f(s)/2\le u
\le t+l-f(s)/2.
\]
Thus setting $Y_{i,j} = P({\cal B}_{l_n}(il_n,jl_n))$,
straightforward computations lead to the following 
representation of discrete process $\omega_{l_n}(kl_n)$: 
\be
\label{sum}
\omega_{l_n}(kl_n) = \sum_{i=-\infty}^{k} \sum_{j=k}^{+\infty} Y_{i,j},
\end{equation}
where $\{Y_{i,j}\}_{i,j}$ are independent infinitely divisible random
variable which satisfies
\be
E\left( e^{pY_{i,j}} \right) = e^{\psi(p)\rho_{i,j}},
\end{equation}
with
\be
\rho_{i,j} = \int_{s\ge l_n} ds/s^2 \int dt H_{i,j,s}(t),
\end{equation}
where $H_{i,j,s}(t)$ is the indicator function
\be
H_{i,j,s}(t) = \mathbb{I}_{[a_{i,j}(s),b_{i,j}(s)]}(t),
\end{equation}
with
\be
a_{i,j}(s) = \max \left((i-1)l_n+f(s)/2,jl_n-f(s)/2 \right),
\end{equation}
and
\be
b_{i,j}(s) = \min \left( il_n+f(s)/2,(j+1)l_n-f(s)/2 \right).
\end{equation}
Let us note that if the function $f$ is bounded (which is the case if we are
under the hypothesis
(\ref{fexact}), i.e., in the case of ``exact'' scaling), the number of terms
in Eq.~(\ref{sum}) is finite.

\begin{figure}[t]
\begin{center}
\includegraphics[width=8cm]{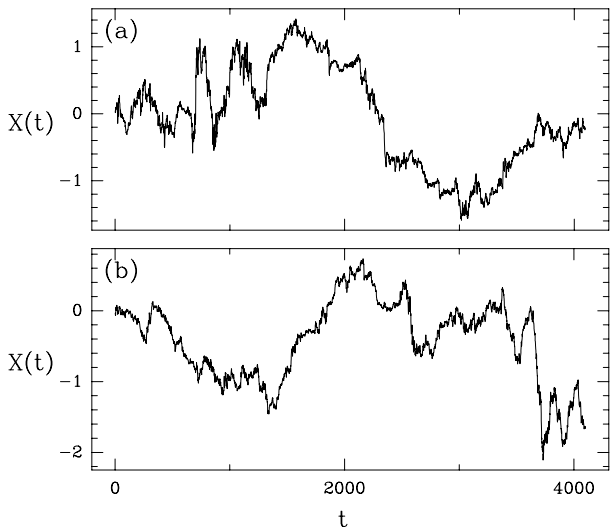}
\end{center}
\caption{MRW signals sampled at rate 1. 
(a) log-Normal MRW with $T=512$ and $\lambda^2=0.05$.
(b) log-Poisson sample with $T=512$, $\gamma = 4$ and $\gamma \ln(\delta)^2 = 0.05$.
In both signals, the cut-off $l$ as been fixed to $1/8$ for
the numerical synthesis.}
\end{figure}

\begin{figure}[h]
\begin{center}
\includegraphics[width=8cm]{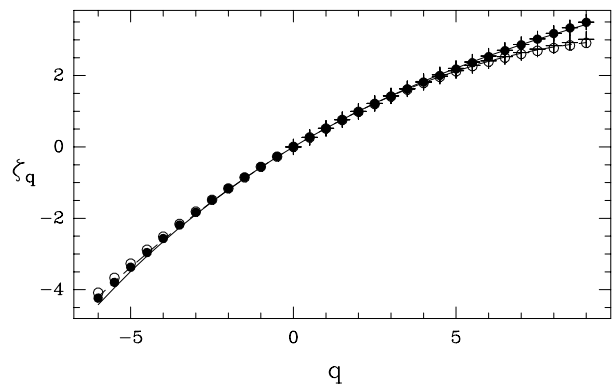}
\end{center}
\caption{$\zeta_q$ spectrum estimation for log-Normal and log-Poisson
MRW's. The exponents have been directly estimated from linear
regression of increment order $q$ absolute moments in doubly
logarithmic representations. These moments have been estimated
using a statistical sample of 256 signals of 64 integral scales.
The symbols ($\bullet$) correspond to log-Poisson estimates while
($\circ$) correspond to log-Normal estimates.
The solid line represents the 
log-Poisson analytical spectrum (\ref{lpzeta}) and
the dashed line to log-Normal analytical spectrum (\ref{lnzeta})
rescaled using Eq.~(\ref{zetaXu}).   
The parameters of both processes are those of 
Fig. 2 and have been chosen so that the
so called intermittency parameter $-\zeta''(0)$, is the same ($0.05$).}
\end{figure}

In the Gaussian case, the situation is simpler because $\omega_l$
is completely characterized by its covariance function,
$\rho_{l}(\tau)$. 
In that case, the 2D synthesis problem can be easily transposed as
a 1D filtering procedure 
by finding a filter $\phi_{l}(t)$ such that
\begin{equation}
\label{phi2}
  \phi_{l} \ast \phi_{l} = \rho_{l} \; ,
\end{equation} 
where $\ast$ stands for the convolution product.
The process
\begin{equation}
   \omega'_{l}(t)= \int \phi_l(t-t') W(dt')
\end{equation}
(where $W$ is here a 1D Wiener noise) will be thus identical
to the normal process $\omega_{l}(t)$ as defined in
Eq.~(\ref{defomega}).
In the original study of Refs. \cite{mrwepj,mrwpre} the 
MRW has been defined along this method.

\subsection{Numerical examples}
In Fig. 3 are shown two samples
of MRW which are respectively log-Normal and log-Poisson
(see  Eqs. (\ref{lnzeta},\ref{lpzeta})).
In both cases we have chosen $T = 512$ sample units, $H=1/2$
($\epsilon$ is a Gaussian white noise). For the log-Normal process, 
$\lambda^2 = 0.05$  
while $\gamma = 4$ and $\delta = e^{-\sqrt{0.05/\gamma}}$ for
the log-Poisson process.
The log-Poisson process has been synthetized using Eq. (\ref{sum}) 
while a simple filtering method was used for the log-Normal process.
Eq. (\ref{phi2}) was solved numerically in the Fourier domain.
In Fig. 4 are plotted the $\zeta_q$ functions
estimated for both processes. These functions have been
obtained from the scaling of the moments estimated
using 256 MRW trials of 64 integral scales long.
The $\zeta_q$ values for negative $q$ have been obtained
using the so-called WTMM method that is a
wavelet based method introduced to study multifractal functions \cite{wtmm1,wtmm2,wtmm3}.
The superimposed analytical formulae obtained
with Eqs. (\ref{zetaXu}), (\ref{lnzeta}) and
(\ref{lpzeta}), fit very well with statitical estimates except
a large negative $q$ values for the log-Normal case.
This can be explained as a finite statistics effect.

\section{Connected approaches}
In this section we review some specific results
concerning respectively log-normal 
and log-Poisson compound MRW.
\subsection{Log-normal MRW}
The log-normal MRW has been originally defined
in Refs. \cite{mrwpre,mrwepj}.
It corresponds to the simplest situation
when the Levy measure has only a Gaussian component.
In that case, the Gaussian process $\omega_l(t)$
can be directly constructed from a 1D white noise, 
without any reference to 2D conical domains. 
This model is interesting because its
multifractal properties are described by
only two parameters, the integral scale $T$ 
and the so-called intermittency parameter
$\lambda^2$.
Some simple estimators of these quantities have
been proposed in \cite{mrwepj}.
Moreover, many exact analytical expressions
can be obtained and notably the
value of the prefactor $K_q$ in Eq.~(\ref{zetaid}).
In \cite{mrwphya}, it is shown that this prefactor
can be written as a Selberg integral \cite{selberg}.
Its analytical expression reads:
\be
\label{Kq}
K_q = T^{q} \prod_{k=0}^{q-1}
\frac{\Gamma(1-2\lambda^2k)^2\Gamma(1-2\lambda^2(k+1))}{\Gamma(2-2\lambda^2(q+k-1))\Gamma(1-2\lambda^2)}.
\end{equation}
It is easy to check that $K_q$ is defined only if
$q < q_{\ast} = 2/\lambda^2$.
We recover the finite moment condition (\ref{infmom}), $q<q_\ast$ with
$\zeta_{q_\ast}=1$.
Notice that, as emphasized by Frisch \cite{frisch}, the existence
of infinite moments can be a drawback of a log-normal
multifractal as a model for experimental situations like
turbulence.
However, for a typical value $\lambda^2 = 5. 10^{-2}$,
$q_\ast = 40$, so a log-normal approximation can be very
good in a range of $q$ values far beyond the limit
associated with the finite size of experimental samples. 
Let us finally mention that the log-normal MRW
can be naturally generalized to a ``multivariate
multifractal model'' which is a multifractal vector
of processes
characterized by an intermittency matrix $\Lambda_{ij}$ \cite{mrwepj}.
This notion of ``joint multifractality" can be very interesting
in many applications. 
\subsection{Multifractal products of cylindrical pulses}
\label{barralmandelbrot}
In reference \cite{barral}, Mandelbrot and Barral introduced
a positive multifractal random measure using products of positive random variables
associated with Poisson points within 2D conical domains.
Their construction is a particular case of MRM. 
It actually reduces to the case where
$\nu(dx)$ is a Levy measure satisfying $\int \nu(dx) x^{-2} < \infty$
(see section \ref{examples}) and where the set 
$\A_l(t)$ is delimited by the function 
$f(l)$ as defined by Eq.~(\ref{fpasexact}).
Let us note that, since they have not considered the full 
domain $\A_l(t)$ associated with 
the function $f(l)$ as defined by Eq.~(\ref{fexact}), 
as explained in section
\ref{asymptscaling}, 
this construction performs asymptotic scaling (\ref{ascaling}) 
but not exact scaling (\ref{escaling}).
However, these authors did not study the scaling
properties of the random measures. They rather focused 
on the pathwise regularity properties.
More precisely, they proved the validity
of the so-called ``multifractal formalism'' 
(see e.g. \cite{wtmm1,wtmm2,wtmm3,fp,jaff}) that relates
the function $\zeta_q$ to the singularity
spectrum $D(h)$ associated with (almost) all realisations
of the process. 
For a given path of the increasing process associated with an MRM,
$D(h)$ is defined as the Hausdorff dimension of
the set of  ``iso-regularity'' points, i.e., the points
where the (H\"older) regularity is $h$.
Barral and Mandelbrot proved that $D(h)$ and
$\zeta_q$ are related by a Legendre transformation. 
Since we proved that $\zeta_q$ is the 
scaling exponent of MRM moments,
it follows that one can estimate the 
singularity spectrum of the MRM paths in the
case of log-Poisson compound statistics.
It should be interesting to extend the Mandelbrot-Barral
theorem to the general log-infinitely divisible MRM and MRW paths.

\section{Conclusion and prospects}

\subsection{Summary and open questions}
In this paper we have constructed a class of
stationary continuous time 
stochastic measures and
random processes that have exact or asymptotic 
multifractal scaling properties in the sense
of Eqs. (\ref{escaling}) and (\ref{ascaling}). 
We have shown how stochastic integration
of an infinitely divisible noise over cone-like
domains, as originally proposed in \cite{schmitt}, 
naturally arises when one wants to ``interpolate''
discrete multiplicative cascades over a continuous range 
of scales within a construction that is invariant
by time translations. The exponential of these
stochastic integrals ($e^{\omega_l(t)}$)
can be interpreted as a ``continuous product''
from coarse to fine scales and thus
as the continuous extension of the multiplicative rule
involved in the definition of discete cascades.
We have shown that
the probalility density functions associated
with MRM and MRW processes 
can have, like discrete cascades, 
fat tails with arbitrary large exponents.
However, unlike their discrete analog, our ``continuous cascades'' 
have stationary fluctuations,  
do not involve any particular scale ratio
and can be defined in a causal way.
This ``sequential'' formulation, as opposed to the 
classical ``top to bottom'' definition of multifractals,
can be very interesting for modelling dynamical processes
(see the next section).
Let us note that we focused in this study on
1D processes but our construction can 
easily be extended to higher dimensions.

It is well known (see Refs. \cite{mrwepj,mrwpre})
that the multiscaling (\ref{escaling}) or (\ref{ascaling})
with a non-linear convex
$\zeta_q$ function cannot extend
over an unbounded range of scales and there
necessarily exists an ``integral scale'' $T$ above
which the scaling of the moments changes.
The existence of such an integral scale can be found
in the general shape of 
function $f(l)$ as discussed in section \ref{asymptscaling}:
One must have $f(l) \sim l$ when $l \rightarrow 0$ 
and $f(l) = o(l)$ when $l \rightarrow +\infty$, so the scale $T$
is a scale that separates these two asymptotic regimes. 
It is remarkable that there exists a particular
expression for the function $f(l)$ (Eq. (\ref{fexact}))
for which the moments in the multifractal regime $l \leq T$ 
satisfy an {\em exact} scaling. The existence of processes
with such properties was not a priori obvious.
In the same section, we have shown that $f(l) \sim l$ is a necessary
and sufficient condition for the  existence of a limit multifractal
object.  From a fundamental point of view, one important
question concerns the unicity of our construction: 
Is any process satisfying (\ref{escaling})
can be represented within the framework we have introduced ?

It remains many open mathematical problems related to the processes
which we introduced in this paper.
Some of them have been already mentionned, notably the questions
related to the construction of stochastic integrals in section
\ref{longrangemrw}. 
As discussed in section \ref{barralmandelbrot},
it should be interesting to generalize the results of Ref. \cite{barral}
in order to link scaling properties and pathwise regularity 
within a multifractal formalism.
Another interesting problem concerns the study of
limit probability distributions associated with MRM for which
fery few features are known.
Like infinitely divisible laws, 
they appear to be related to some semi-group structure.
Finally, one can wonder if log-infinitely processes
we have defined are not the natural candidates to be described
within the framework of ``Markovian continuous cascades''
as introduced in Refs. \cite{mark1,mark2}.

\subsection{Possible applications}
One of the main issues of the present work was to
construct a wide family of multifractal processes (or measures)
that are likely to be pertinent models in many  
fields were multiscaling laws are observed.
Naturally, the first application of which one can
think, is fully developed turbulence. Turbulence and multifractals
share a long history and we refer the reader to Ref. \cite{frisch}
for a review on the ``intermittent'' nature of turbulent fields.
Recently, new aspects of turbulence were studied
by considering fluid dynamics from a Lagrangian point of
view. This was possible because two groups
developed new experimental devices based 
on a fast imaging system \cite{boden} or
ultrasound techniques \cite{pinton1,mordant} allowing
for a direct measurement of the velocity of a single tracer
in a turbulent flow. In a recent works, Pinton and his 
collaborators \cite{pinton1,mordant} studied the
intermittency of Lagrangian trajectories and
related it to the slow (logarithmic) decay of the
particle acceleration correlations \cite{mordant,delour}, 
very much like for a MRW model.
In other words, these authors found that the
turbulent Lagrangian dynamics is very well described
by an equation of Langevin type with a driving force
amplitude similar to $e^{\omega_l(t)}$
involved in the MRW definition. 
The understanding of the physical origin of such
dynamical correlations and the link between
Lagrangian and Eulerian statistics is 
a very promising path towards the explanation
of the intermittency phenomenon in fully developed turbulence.

Besides turbulence, ``econophysics" \cite{bBou99,bMan00,aFar99}
is an emerging field where fractal and multifractal concepts
have proven to be fruitful. 
Indeed, many recent studies
brought empirical evidences for the multifractal
nature of the fluctuations of financial markets (see \cite{mrwepj}
and reference therein). 
Some physicists raised an interesting analogy between
turbulence and finance \cite{bMan00,aGha96,aArn98,schmittfin}.
Logarithmic decaying correlations and ``1/f'' power spectrum 
have been directly 
observed for various time series \cite{aArn98,aBon99}, so
it is reasonable to think that MRW models are
well suited for modeling financial time series \cite{mrwepj}. 
The versatility of infinitely divisible MRW
is very interesting to account for various stylized facts
of financial times series such that the multiscaling,
the power-law tail behavior of return pdf and therefore
such models can be very helpful for financial engineering
and risk management.

Among all the remaining disciplines where multifractality has been
observed, one can mention 
the study of network traffic \cite{taqqu2,abry,gilbert},
geophysics and climatology \cite{waymire,tessier,clouds}, 
biomedical engineering \cite{heart}, the modelling of natural images \cite{turiel},... On a more theoretical ground, ``continuous branching trees'' and
log-correlated random processes have also been considered
in the physics of disordered systems \cite{derrida, carpentier}.
The study of free energy density in the thermynamic limit in presence
of log-correlated disorder (or equivalently on topologies involving
random disordered trees) raise questions very similar to
the study of limit MRM addressed in this paper. One can hope
that pushing forward this analogy in a promising
prospect to get significant results in both fields.

\begin{acknowledgments}
We would like to thank Jean Delour for valuable discussions and
Jean-Fran\c{c}ois Pinton for providing us his latest
results on Lagrangian turbulence.
\end{acknowledgments}

\appendix
\section{Mean square convergence of a MRM}
Let $M_l(dt)$ defined as in Eq.~(\ref{Ml}).
In this section we prove that, assuming $\psi(2)<1$, one has
\be 
\label{msm}
  M_l([0,t]) \operatornamewithlimits{\longrightarrow}_{l \rightarrow 0}^{m.s.} M([0,t])
\end{equation}
Let us define
\be
R_{l,l'}(\tau) = E\left( e^{\omega_l(u)+\omega_{l'}(u+\tau)} \right) \; .
\end{equation}
In order to prove (\ref{msm}), let us first show that, if $l' \leq l$:
\be
\label{rl}
  R_{l,l'}(\tau)  = R_{l,l}(\tau) = e^{\rho_l(\tau) \psi(2)}
\end{equation}
with $\rho_{l}(\tau)$ as defined as in (\ref{defrho}) and
(\ref{rhoexact}).

The first equality in Eqs.~(\ref{rl}) comes directly from the 
assumption $\psi(1) = 0$ while the second equality is a 
particular case of the identity (\ref{char}), where $q=2$, $p_1 = p_2 = -i$
and $t_1 -t_2 = \tau$.

Let us  show, that,
$\forall \; \epsilon$, $\exists \; l_0$, $\forall \; l,l' < l_0$,
$E \left[ (M_l(t)-M_{l'}(t))^2 \right] < \epsilon$.
Let us suppose that $l'\leq l$.
Then,
\begin{eqnarray*}
&& C(l,l',t)  =  E \left[ (M_l([0,t])-M_{l'}([0,t]))^2 \right]   \\
&& =  E[M_l^2([0,t])]+E[M_{l'}^2([0,t])]
-2 E[ M_l([0,t]) M_{l'}([0,t])]  \\
&  & =  \int_0^t \int_{0}^{t} \left( E\left(e^{\omega_l(u)+\omega_l(v)}\right)+
E\left(e^{\omega_{l'}(u)+\omega_{l'}(v)}\right) \right) du dv \\
& & -  2 \int_0^t \int_{0}^{t} E\left(e^{\omega_l(u)+\omega_{l'}(v)}\right) du dv
\end{eqnarray*}
Thus, thanks to Eq.~(\ref{rl}), we get, after a little algebra
\begin{eqnarray*}
C(l,l',t) & \le &  
D t  \int_{0}^{t} \left( R_{l',l'}(u)-R_{l,l}(u) \right) du, \\ 
& \le &  D t  \int_{0}^{l} R_{l',l'}(u) du, \\ 
& \le &  D t  \int_{0}^{l'} R_{l',l'}(u) du + D t\int_{l'}^{l} R_{l',l'}(u) du \\
& \le &  E t {l'}^{1-\psi(2)} + E t  \left( l^{1-\psi(2)} - {l'}^{1-\psi(2)} \right).
\end{eqnarray*}
where $D$ and $E$ are positive constants. 
Since $\psi(2) < 1$, we see that $M_l([0,t])$ is a 
Cauchy sequence and thus converges in mean square sense. \\
The previous computations also prove the  convergence of all finite dimensional vector $\{M_l[t_1,t_1+\tau_1],\ldots,
M_l[t_n,t_n+\tau_n]\}$.
In order to prove the existence of the limit $M([0,t])$ as a stochastic process,
one further needs a ``tightness'' condition. Such a condition
can be obtained along the same line as for previous computation. 
Indeed, provided $\psi(2) < 1$,
one can bound the order $2$ moment of $M_l([0,t])$, $\forall t,l$:
\[
  E \left( M_l([0,t]^2 \right) \leq C t^{2-\psi(2)} \; ,
\]
where $C$ does not depend on $l$.
This achieves the proof.

\section{Mean square convergence of a MRW 
with a fractional Gaussian noise $\epsilon_H$}
In order to simplify the proof and to avoid
technical complications, let us show the mean 
square convergence of the process
\be
 X_l(t) = \int_0^t dW_{H}(t) e^{\omega_l(t)}
\end{equation}
where $dW_H$ is a continuous fGn which covariance is ($H \neq 1/2$)
\be
  \gamma_H(\tau) = \sigma^2 H(2H-1) \tau^{2H-2}
\end{equation}

Rigorously speaking the previous integral is not well defined
but the proof of the convergence of the discrete version (\ref{mrwfmb})
is very similar.
Let us now 
\be 
\label{ms}
  X_l(t) \operatornamewithlimits{\longrightarrow}_{l \rightarrow 0}^{m.s.} X(t)
\end{equation}
provided $H$ satisfies:
\be
\label{condition}
   H > 1/2 + \psi(2)/2
\end{equation}
We proceed along the same line as in Appendix A.
In order to prove (\ref{ms}), we have to prove, that,
$\forall \; l,l' \leq l$,
$E \left[ (X_l(t)-X_{l'}(t))^2 \right] \rightarrow_{l \rightarrow 0} 0 $. Let \be
E = E \left[ (X_l(t)-X_{l'}(t))^2 \right].
\end{equation}
Thanks to Eq.~(\ref{rl}), we have, 
\begin{eqnarray*}
E & = & E[X_l^2(t)]+E[X_{l'}^2(t)]
-2 E[ X_l(t) X_{l'}(t)] \\
 & = &  \int_0^t \int_{0}^{t} \left( E\left(e^{\omega_l(u)+\omega_l(v)}\right)+
E\left(e^{\omega_{l'}(u)+\omega_{l'}(v)}\right) \right) \gamma_H(|u-v|) du dv \\ 
 & - &  2 \int_0^t \int_{0}^{t} E\left(e^{\omega_l(u)+\omega_{l'}(v)}\right)
 \gamma_H(|u-v|) du dv\\
 & = &  \int_0^t \int_{0}^{t} \left(R_{l'l'}(|u-v|)-R_{l,l}(|u-v|)\right)
\gamma_H(|u-v|) du dv 
\end{eqnarray*}The last integral behavior in the limit $l \rightarrow 0$ can 
be easily evaluated.
After some simple algebra we get:
\[
E \left[ (X_l(t)-X_{l'}(t))^2 \right] = O\left( l^{2H-\psi(2)-1} \right)
\]
Thus if condition (\ref{condition}) is satisfied,
i.e. $2H-\psi(2)-1 > 0$,
$X_l(t)$ is a 
Cauchy sequence and thus converges in mean square sense.

\end{document}